\documentstyle[aps,12pt]{revtex}

\begin{document}

\preprint{EFUAZ FT-99-71}

\title{Do Zero-Energy Solutions of Maxwell
Equations Have the Physical Origin Suggested by A. E. Chubykalo?
[Comment on the paper in Mod. Phys. Lett. A13 (1998) 2139-2146]
\thanks{Submitted to ``Modern Physics Letters A"}}

\author{{\bf Valeri V. Dvoeglazov}}

\address{Escuela de F\'{\i}sica, Universidad Aut\'onoma de Zacatecas\\
Apartado Postal C-580, Zacatecas 98068 Zac., M\'exico\\
E-mail: valeri@ahobon.reduaz.mx\\
URL: http://ahobon.reduaz.mx/\~~valeri/valeri.htm
}

\date{August 17, 1999}

\maketitle

\medskip

\begin{abstract}
Zero-energy solutions of Maxwell's equations have been discovered and
re-discovered many times in this century. In this comment
we show that the paper by A. E. Chubykalo (which, in fact, also comments
on the recent papers) did not explain  physical interpretation of
this solution and it used doubtful (up-to-date) postulates.
Different explanations of the considered problem are possible.
\end{abstract}

\bigskip
\bigskip

The dynamical free-space Maxwell equations can be written in the form
\begin{mathletters}
\begin{eqnarray}
{\bbox\nabla} \times [{\bf E} -i{\bf B}] + i (\partial/\partial t)
[{\bf E} - i{\bf B}] &=& 0\, ,\\
{\bbox\nabla} \times [{\bf E} +i{\bf B}] - i (\partial/\partial t)
[{\bf E} + i{\bf B}] &=& 0 \, .
\end{eqnarray}
\end{mathletters}
(cf., e.g., formulas (4.19)-(4.22) in ref.~\cite{Wein}).
The paper~\cite{Chub} begins with the statement that this set
support {\it non-trivial} solutions with energy $p_0=\pm \vert {\bf
p}\vert$ and $p_0 =0$, and refers in this connection to the
works~\cite{Opp,Ahl}, where the equations (1) have been
written for {\it 3-vectors} $\phi_{_{L,R}} ({\bf p})$  as opposed
to the {\it fields} in (1a,b).\footnote{The signs in the formulas (1) of
ref.~[6a], (1,2) of ref.~\cite{Chub} and (20,21) of ref.~[6b] are
interchanged. No explanations present therein.} Then, trying to explain
the negative-energy and zero-energy solutions Chubykalo obtained a new
form of the energy density of the electromagnetic
field~\cite[Eq.(28)]{Chub}. It is based on his
instantaneous-action-at-a-distance electrodynamics which includes {\it
implicit} dependence of the fields on  time.

In the present note we analyze his work and show that it does not
contain sound physical explanation. The main postulate
is also required solid experimental and mathematical basis.
Nevertheless, we agree that the question of {\it non-transverse}
solutions~\cite{Ev,Dvor2,Espos,Dvon1,Dvon2,Rodrig,book} requires  further
research.

First of all, I want to mention that the so-called Oppenheimer-Ahluwalia
$E=0$ solutions\footnote{Here we use the terminology of the commented
paper.} was discovered by many researchers, whose
works~\cite{Opp,Good,Gian,Ahl,Dvo,Ev} should be paid credits
to.\footnote{Beforehand, I apologize before those discoverers of the
non-transverse solutions of the Maxwell's equations whom I did not
mention here.  I am sure that a lot of papers remains
unnoticed.}$^,$\footnote{Very unfortunately, neither Ahluwalia
(the first papers are in ref.~\cite{Ahl}) nor Chubykalo papers contain
sufficient list of references. I informed Dr. Ahluwalia about previous
related works as earlier as in 1994-95.} They have been analyzed further
in~\cite{Dvor1,Dvor2,page}.\footnote{These papers proved that $E=0$
solution is not a varying solution in time and it may be {\it unbounded}.}
A solution of the similar nature in two-particle system was also
considered, see, for example, ref.~\cite{Mosh}.

A. Chubykalo~\cite{Chub} considers Landau and Lifshitz~\cite[\S
31]{Lan} derivation of the Poynting theorem in the Section I. He objects
it on the basis that it does not provide zero energy (negative-energies as
well) unless the fields ${\bf E}$ and ${\bf B}$ are equal to zero.
Furthermore, in the Landau-Lifshitz procedure ``one implicitly neglects
a radiation field which can go off to infinity" (in the opinion of the
author of~\cite{Chub}).\footnote{Indeed,
authors of different textbooks on classical/quantum electrodynamics
(e.~g., ref.~\cite[p.105]{Barut}) also assumed the postulate that the
fields, their derivatives and Lagrangian, all tend to zero at spatial
infinity.} Chubykalo then divides the electromagnetic fields into {\it
two} parts (which depend on time explicitly and implicitly), writes
corresponding Maxwell's equations\footnote{In my opinion, equations
(20,21) of ref.~\cite{Chub} are just another form of the Maxwell equations
for this particular case, in the sense that there is no new physical
content if one expects that the Maxwell electrodynamics describes also the
Coulomb interaction.  Brownstein in~\cite{Bron} showed that, in fact,
confusions sometimes arise when physicists use the same symbol
$\partial/\partial x^\mu$ for both partial and total derivative.
Unfortunately, Brownstein did not cite the previous works on the
subject~\cite{Chubo}.} and, in fact, he {\bf postulates} that some part of
the total electromagnetic field (${\bf E}^\ast$ and ${\bf H}^\ast$ in the
notation of~\cite{Chub} to be precise) does {\bf not vanish} at the
spatial infinity! See the formula (26) therein. In such a way he {\bf
re-normalizes} the energy density subtracting the quantity ${{\bf
E}^{\ast\, 2} + {\bf H}^{\ast\, 2} \over 8\pi}$ from the energy density
used by Landau {\it et al.}, see formulas (25)-(28) in~[2]. The Chubykalo
final result for energy density $\omega$ is
\begin{equation} \omega =
\frac{2{\bf E}^\ast\cdot {\bf E}_0 + 2{\bf H}^\ast \cdot {\bf H}_0 + E_0^2
+ H_0^2}{8\pi} = \frac{{\bf E}_{tot}^2 + {\bf H}_{tot}^2}{8\pi} -
\frac{{\bf E}^{\ast\, 2} + {\bf H}^{\ast\, 2}} {8\pi}\, .\label{ne}
\end{equation} The main problem with the
Chubykalo derivation is the following:  the integrals are {\bf divergent}
when they extend over all the space.  This refers to the first term of
(25), the integrals in (26) and the dot products in the numerator of the
first term of (27).  This is easily proven on observing that if ${\bf
E}^\ast$ and ${\bf H}^\ast$ do not tend to zero at infinity, they have the
asymptotic behavior of $O(r^\alpha)$, $\alpha \geq 0$. In the mean time,
${\bf E}_0$ (for instance) refers to Coulomb (Coulomb-like)
electromagnetic field and, thus, has the asymptotic behavior $O(1/r^2)$.
The factor $r^2$ in the volume element cancels $1/r^2$, leading to the
total asymptotic of the integrated expression $O(1), r \rightarrow
\infty$ in the better case.  I am not going to discuss the mathematical
validity of subtracting infinite expressions in this case, leaving this
problem to specialists. But, to the best of my knowledge, some persons
claimed that such procedures do not have sound mathematical
foundation~\cite{Dirac}. Furthermore, even if one accepts its validity the
total energy resulting from integration of (28) over the whole space is to
be infinite!?  It was noted by Barut~\cite[p.105]{Barut} that in the case
of non-vanishing fields at the spatial infinity {\bf we cannot
expect to find globally conserved quantities}.

In the mean time, the definitional problems of the 4-momentum
of the electromagnetic field and
the problem of the compatibility of the energy-momentum definition
and the Poynting theorem (the formula (10) of~\cite{Chub})
with the Relativity Theory (SRT)
are the old ones~\cite{Pauli,previous}. Pauli~\cite{Pauli}  stated: ``We
therefore see that {\it the Maxwell-Lorentz electrodynamics is quite
incompatible with the existence of charges, unless it is supplemented by
extraneous theoretical concepts.}" But, Rohrlich~\cite{Rohr}
and Butler~\cite{Butler}
seem to have solved this problem establishing for the
steady state (when  ${\bf B} = {\bf v}\times {\bf E}$ is valid) the new
expressions for the energy and momentum of the electromagnetic system
\begin{mathletters}
\begin{eqnarray}
P^0 &=& \gamma \int \left [({\bf E}^2 -{\bf B}^2)/8\pi \right ] d^3
\sigma = \gamma m, \\
{\bf P} &=& \gamma {\bf v} \int \left [({\bf E}^2 -{\bf B}^2)/8\pi \right
] d^3 \sigma = \gamma {\bf v} m,
\end{eqnarray}
\end{mathletters}
(equations (3.39) and (3.40) of~[26b]); $m$ is the
electromagnetic mass and $\gamma$ is the usual contraction factor in the
SRT.  Butler wrote: ``...In the derivation of the Poynting's theorem {\it
only the two curl equations were used}. The two divergence equations were
ignored. Therefore Poynting's theorem is covariant only if both divergence
terms are equal to zero!...  only in the absence of charge! That is the
reason that Pauli wrote..." Rohrlich indicated that the equations (3.39)
and (3.40) are deduced from more general form (see (3.23) and (3.24)
therein).  The old definitions $U = {1\over 8\pi} ({\bf E}^2 +{\bf
B}^2)$ and ${\bf S} = {1\over 4\pi} ({\bf E}\times {\bf B})$ are valid
only when ${\bf v}\equiv 0$ (in fact this signifies that the old
definitions are erroneous when applied to the electrodynamics with
moving charges).\footnote{See also the first paragraph on p. 1313
in~[26b].} Unfortunately, the expression of Chubykalo~\cite[Eq.(28)]{Chub}
seems not to reduce to the Rohrlich-Butler formula in the steady-state
limit (i.e.  when ${\bf E}^\ast = {\bf H}^\ast =0$).

Next, we shall try to deepen understanding the origins why
did Chubykalo obtain such a strange result. We shall pay attention to the
Lagrangian formalism~\cite{Barut,Bogol}.
In general, it is not clear if there may exist a situation when
the fields does {\it not} tend to zero at the spatial infinity,
but their variation (and the variation of derivatives) do (the latter is
necessary for the validity of the Lagrangian formalism).  But, it is
obvious that in the Chubykalo case one may wish to use the variation
procedure for {\it two} types of potentials in order to try to obtain the
formula (\ref{ne}).  The ordinary Lagrangian \begin{equation} {\cal L}
=-{1\over 2} {\partial A_\mu \over \partial x^\nu} {\partial A^\mu \over
\partial x_\nu} \end{equation} indispensably leads in the radiation gauge
$A_0 =0$ to the standard expression for the
Hamiltonian:\cite[p.147]{Ryder} \begin{equation} {\cal H} = {1\over 2}
\int (\vert {\bf B} \vert^2 + \vert {\bf E} \vert^2) d^3 x = {1\over 2}
\int \left [ \vert \nabla \times {\bf A}\vert^2 +{1\over c} (\partial {\bf
A}/ \partial t)^2 \right ] d^3 x \, .
\end{equation}
The first way to
think about the generalization of the Lagrangian for the Chubykalo case is
to try to add the total 4-derivative\footnote{The Barut formula in
ref.~\cite[p.107]{Barut} (see the formula next to (3.62)) seems to be
incorrect due to $\partial_\nu \Gamma^\nu$ may depend on
the field derivatives and contribute to the second term of (3.62).}
But, in seemingly appropriate case $\Gamma^\mu = A^\nu F^{\mu}_{\quad\nu}$
we come to the situation when the second-order derivatives
enter into the Lagrangian and the Ostrogradsky procedure
becomes to be necessary.

Here, we suggest to generalize the Lagrangian for electromagnetic
field described by the 4-vector potential in the following
way:\footnote{We denote the complex conjugation
by a line over a letter. This is used in order not to confuse
with the ``star" notation for the free fields ${\bf E}^\ast$ and ${\bf
B}^\ast$ in~\cite{Chub}.} \begin{equation} {\cal L} = - {1\over 4}
{\partial {\cal A}_\mu \over \partial x^\nu } {\partial {\cal A}^{\mu}
\over \partial x_\nu } - {1\over 4} {\partial \overline{{\cal A}}_\mu
\over \partial x^\nu } {\partial \overline{{\cal A}}^{\mu} \over \partial
x_\nu }\, .\label{nel} \end{equation} Thus, we introduce the {\it complex}
4-potential ${\cal A}_\mu = A_\mu^{(elm)} + i A_\mu^\ast$, where
$A_\mu^\ast $ corresponds to the fields ${\bf E}^\ast$ and ${\bf H}^\ast $
which do not vanish at the infinity. Obviously, it is Hermitian and of the
first order in derivatives. If one varies  over $A_\mu^{(elm)} $, the
common-used potential,  and $A_\mu^\ast $ independently (what, in fact,
the Chubykalo-Smirnov-Rueda papers assumed), the new form of the
Lagrangian (\ref{nel})
\begin{equation} {\cal L} =-{1\over 2} {\partial
A_\mu^{(elm)} \over \partial x^\nu} {\partial A^{\mu\,(elm)} \over
\partial x_\nu} + {1\over 2} {\partial A_\mu^{\ast} \over \partial x^\nu}
{\partial A^{\mu\,\ast} \over \partial x_\nu}
\end{equation}
leads to Eq. (\ref{ne}).
So, this suggestion of ours may have some connections with
ref.~\cite{Dvon2}, where the idea of the {\it complex} 4-vector potential
has been proposed. This idea should be compared with the Stepanovski\u{\i}
work~\cite[p.189]{Stepan}, which speculates on the unification of gravity
with electromagnetism on the basis of complex
potentials in the framework of the Weyl theory.~\footnote{The referee
of {\it Foundation of Physics Letters} rejected [15a] on the basis of
no-proof (in his opinion) of the idea of complex potentials. I disagreed
with him on the basis of several papers exploring 1) possibility of
various field operators (e.g., the negative energy part of a field
operator can be not $C$ conjugate but $CP$ conjugate as in the case of
neutrinos -- other neutral particles); 2) complex-space coordinates; 3)
possibility of presenting the spinorial affine connection as a complex
4-vector~\cite{Weyl,Stepan}.  Furthermore, the concurrent paper~[15b] has
been published. I consider this referee to be biased and directed to the
destruction of my work.}

Finally, I want to recall about the problem of energy in {\it quantum}
electrodynamics. In ref.~\cite[p. 89]{Bogol} the book authors
write {\it explicitly}: ``{\it Quantization (12.2) [the standard one]
does not provide the positive definiteness of the mean value of energy}"
(sic!) After Gupta and Bleuler~\cite{indef} by an artificial
procedure (``by hands") they 1) assumed zero component of
the 4-potential to be {\bf anti-Hermitian} (sic!),\footnote{This
has again relevance to the idea of the complex
4-potential~[15].} 2)
indefinite metrics in the state-vectors space, and 3) the weak Lorentz
condition.  As a result, on the basis of the above postulates, they
obtained the positive definiteness of the energy. If one wants
to have negative and zero energies in the quantization of the
electromagnetic field (as Chubykalo and Ahluwalia) the answer is simple:
they should not apply additional postulates given few lines
above.\footnote{Seems, the referee of {\it Foundation of Physics Letters}
did not read the University textbooks.}

Finally, we explained the physical and mathematical content of the
paper~\cite{Chub}. The conclusion is: {\it we are not yet convinced
in the necessity  of correction of the formula for the energy density for
radiation field because of the absence of firm experimental and
mathematical bases in~[2].}

\acknowledgments
I greatly appreciate discussions
with Profs.  A.  Chubykalo (1995-1999), patient answers on my questions by
Profs. Y. S.  Kim, M. Moshinsky and Yu. F. Smirnov, and the useful
information from Profs. A. F. Pashkov (1982-99) and E. Recami (1995-99).
Frank discussions with Prof. D.  Ahluwalia (1993-98) are acknowledged,
even if I do {\it not}  always agree with him.

Zacatecas University, M\'exico, is thanked for awarding
the full professorship.  This work has been partly supported by
the Mexican Sistema Nacional de Investigadores and the Programa
de Apoyo a la Carrera Docente.

\end{document}